\documentclass[aps,prl,groupedaddress, amsmath,amssymb]{revtex4}
\usepackage{graphicx}
\usepackage{float}
\usepackage{epstopdf}
\usepackage{natbib}
\usepackage{hyperref}
\hypersetup{
   bookmarks=false,
   colorlinks,
   citecolor=blue,
   filecolor=blue,
   linkcolor=black,
   urlcolor=blue
}

\usepackage[utf8]{inputenc}
\usepackage[T1]{fontenc}

\begin{document}


\title{Fiber cavities with integrated mode matching optics}

\author{Gurpreet Kaur Gulati}
\altaffiliation{Department of Physics and Astronomy, University of Sussex, Falmer, BN1 9QH, U.K.}
\author{Hiroki Takahashi}
\altaffiliation{Department of Physics and Astronomy, University of Sussex, Falmer, BN1 9QH, U.K.}
\author{Nina Podoliak}
\altaffiliation{Optoelectronics Research Centre, University of Southampton, Southampton SO17 1BJ, U.K.}
\author{Peter Horak}
\altaffiliation{Optoelectronics Research Centre, University of Southampton, Southampton SO17 1BJ, U.K.}
\author{Matthias Keller}
\altaffiliation{Department of Physics and Astronomy, University of Sussex, Falmer, BN1 9QH, U.K.}

\date{\today}
\begin{abstract}
In fiber based Fabry-P\'{e}rot Cavities (FFPCs), limited spatial mode matching between the cavity mode and input/output modes has been the main hindrance for many applications. We have demonstrated a versatile mode matching method for FFPCs. Our novel design employs an assembly of a graded-index and large core multimode fiber directly spliced to a single mode fiber. This all-fiber assembly transforms the propagating mode of the single mode fiber to match with the mode of a FFPC. As a result, we have measured a mode matching of 90\% for a cavity length of $\sim$400 $\mu m$. This is a significant improvement compared to conventional FFPCs coupled with just a single mode fiber, especially at long cavity lengths.  Adjusting the parameters of the assembly, the fundamental cavity mode can be matched with the mode of almost any single mode fiber, making this approach highly versatile and integrable.
\end{abstract}

\maketitle
Optical Fabry-P\'{e}rot cavities are essential tools for a wide range of applications in science and technology. In recent years, newly developed laser ablation techniques made it possible to fabricate high-quality mirrors directly on the facets of optical fibers, leading to the advent of  miniaturized fiber-based Fabry-P\'{e}rot cavities (FFPCs) \cite{Hunger2012}. FFPCs support significantly smaller mode volumes compared to the conventional Fabry-P\'{e}rot cavities based on bulk optical mirrors. The resulting quantum mechanical effects have been exploited in atomic \cite{Colombe2007,Steiner2013}, molecular \cite{Tonielli2010}, solid state \cite{Albrecht2013,Miguel2013,Besga2015} and optomechanical systems \cite{Flowers2016} and are crucial for applications in quantum science and technology. Furthermore, due to their compactness and integration with fiber optics, FFPCs have many potential technological applications beyond fundamental scientific research, e.g., tunable frequency filters \cite{Miller1990}, scanning microscopy \cite{Mader2015}, gas sensing \cite{Petrak2014} and micro-fluidics \cite{Ni2016}.
Recently several novel laser machining methods have been developed \cite{Takahashi2014,Ott2016} and detailed characterizations of FFPCs has been carried out \cite{Bick2016,Gallego2016,Benedikter2015,Podoliak2017}. 
Despite these latest developments, the limited spatial mode matching of FFPCs to their inputs and outputs is still a substantial hindrance for many applications. 
In contrast to conventional Fabry-P\'{e}rot cavities, where mode matching can be achieved with free space optics, this is not possible with FFPCs. Usually, FFPCs are directly coupled to a single mode (SM) fiber where the mode matching is simply the overlap between the cavity mode and the output mode emanating from the SM fiber. There are no controllable elements once the cavity geometry and the type of the SM fiber are fixed.  In particular, the mode matching efficiencies will be severely limited when the cavity length becomes much longer than the Rayleigh range of the output mode of the SM fiber.
Using large mode field diameter fibers such as photonic crystal fibers \cite{Ott2016} or tapered fibers, which adiabatically transform the fiber mode diameter to that of the cavity, the mode matching can be improved. However, the mismatch of the wavefront curvatures ultimately limits the mode matching efficiency.

In this letter we demonstrate a novel design of FFPCs which enables optimized mode matching between the cavity and fiber output mode, and which is also free from the constraints of the cavity and fiber geometries. Using a concatenation of a graded-index (GRIN) and multimode (MM) fibers spliced together, a versatile all-fiber mode-matching assembly can be built. With this assembly directly attached to a FFPC, not only the field diameter of the fiber mode but also its wavefront curvature can be precisely engineered to match the mode of the FFPC. 
As a result, this technique enables the use of almost any optical SM fibers as an input/output for FFPCs in a fully integrated manner. We have designed, built and characterized a FFPC with integrated mode matching assembly and have measured a mode matching efficiency of up to 90\% where the conventional design only achieves $\sim$30\%.

\begin{figure}[t]
  \begin{center}
   \includegraphics{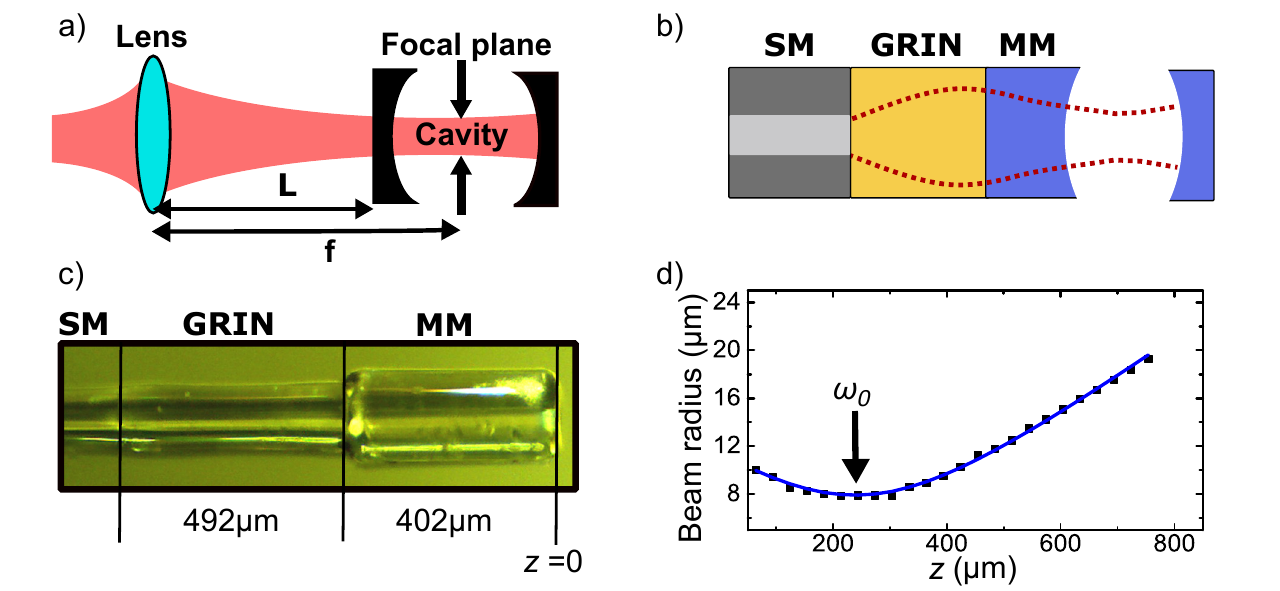}
   \caption{
    a) Mode matching between an optical cavity and a laser beam in free space. b) Design of the mode matching fiber assembly. c) Actual mode matching fiber assembly (FA) with the GRIN  and MM fibers of length 492 $\mu m$ and 402 $\mu m$ respectively. The lengths of the respective fibers determine the waist size and waist position of mode at the output of the FA. d) Mode radius of the output from the FA, where $z$\,=\,0 is the laser ablated end facet of the MM fiber. Waist size ($\omega_{0}$) of 8.1 $\mu m$ at $z$ = 230 $\mu m$ is obtained by fitting the data with the propagation of a Gaussian beam.}
   \label{fig:realpic}
 \end{center}
\end{figure}

In order to achieve an optimal mode overlap between an optical cavity and a laser beam in free space, a single lens with an appropriate focal length ($f$) has to be located at the right distance ($L$) from the position of the cavity (see Fig.~\ref{fig:realpic}~(a)).
Our design follows the same concept but in an all-fiber implementation (Fig.~\ref{fig:realpic}~(b)).
As the focusing element we employ a GRIN fiber (OFS, BF04432-1) with its length determining the effective focal length. Firstly, the GRIN fiber is spliced (Fujikura FSM1000 ARC Master) to a cleaved single mode fiber (IVG, Cu800). 
Then the GRIN fiber is precisely cleaved (Fujikura CT100) under a microscope \cite{Wang2014} to a predefined length to within $\pm 15\,\mu m$. The length of GRIN fiber is adjusted such that the resulting mode waist matches that of the FFPC. Even though the splice losses between the SM and GRIN fibers are low ($\approx$ 0.01\,dB as specified by the splicer analysis), they result in a modification of the effective mode field diameter at the SM-GRIN interface, which needs to be taken into account when cleaving the GRIN fiber. The actual mode waist is confirmed by measuring the mode profile at the output of the SM-GRIN assembly along the axial direction. To this end, a laser beam is coupled into the other end of the SM fiber and the output beam from the GRIN fiber is measured using a knife edge translating across the beam. Fitting the measurement with the propagation of a Gaussian beam, the waist size and its position with respect to the end facet of the GRIN fiber is found. To position the cavity mirrors at the optimal distance from the GRIN fiber a piece of a MM fiber is spliced to it. This fiber serves as a spacer as well as a carrier for the cavity mirror. Due to the large core size of the MM fiber, it does not disturb the free propagation of the mode shaped by the GRIN fiber. Based on the waist position measurement, the MM fiber is cleaved at the optimal position for the desired cavity geometry. 
Finally, the end facet of the MM fiber is laser ablated \cite{Takahashi2014}.
The resulting mode matching fiber assembly (FA), shown in Fig.\ref{fig:realpic}(c), is a concatenation of a SM, GRIN and MM fiber which are directly spliced together, producing a co-axial all-fiber structure. The final mode profile measured at the output of FA, is illustrated in Fig.\ref{fig:realpic}(d) which clearly depicts that the  output beam is converging to a waist outside the assembly. After characterizing the mode, the end facet of the MM fiber is coated with a highly reflective dielectric coating (20 ppm losses at 866\,nm from Laseroptik GmbH). %

To evaluate the mode matching between the fiber and cavity modes, we set up a FFPC. One of the fibers is clamped on a custom-made fiber mount, including a sheer piezo actuator which can translate the fiber along its axis for fine tuning and scanning of the cavity length. This mount is in turn attached to a kinematic prism stage to allow angular adjustment. The second fiber is mounted on a motorized translation stage facilitating 3-dimensional adjustment of the fiber position. In particular, the fiber can be translated to adjust the cavity length. To facilitate the fiber cavity alignment and the measurement of  the distance between the fiber tips, two perpendicularly arranged microscopes are used to radially observe the fiber cavity. Hence, the cavity length can be measured to within $\pm 5\mu m$.
After adjusting the fiber alignment with the microscopes, laser light at $854\,\mathrm{nm}$ is coupled into one of the fibers while the light transmitted through the cavity and entering into the second fiber is detected using a photo diode at the uncoated fiber end. The fibers are then further aligned by observing the cavity transmission, to maximize the transmission of the TEM$_{\small{00}}$ mode, while scanning the length of the cavity with the piezo actuator.

\begin{figure}[t]
  \begin{center}
    \includegraphics[scale=0.75]{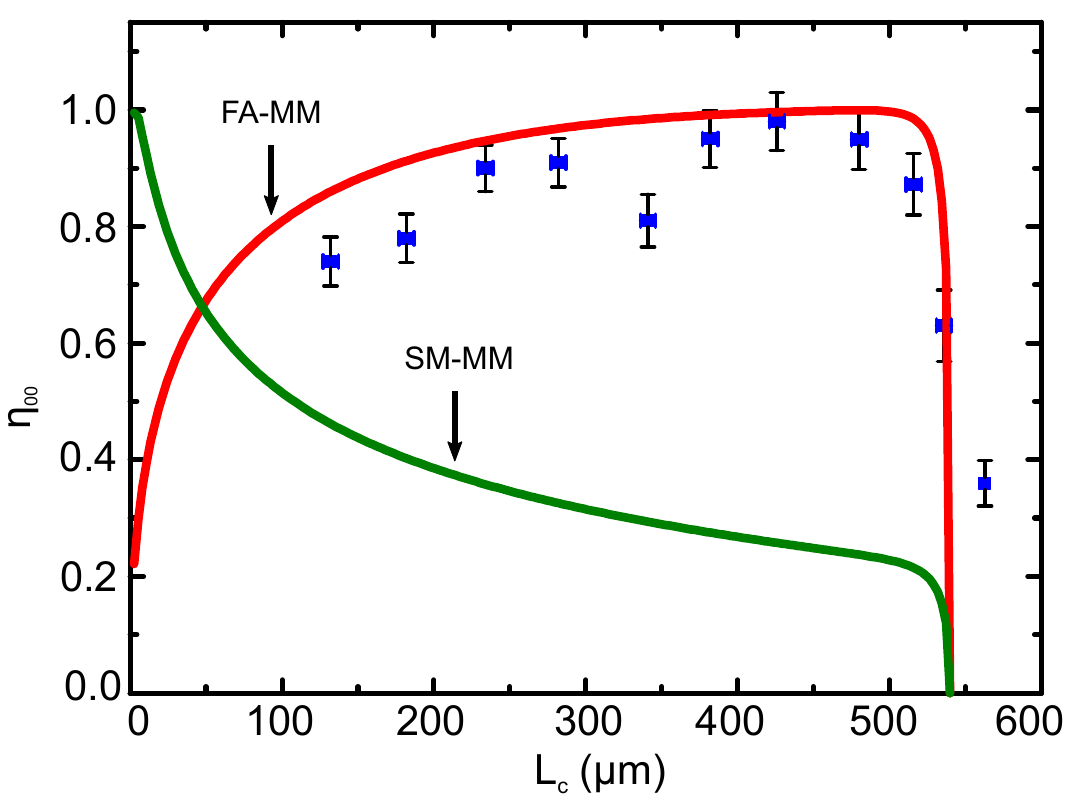}
   \caption{\label{fig:modematch}
 The data points are estimated mode matching efficiencies ($\eta_{00}$)  as a function of cavity length ($L_{c}$), calculated by using Eq.~(\ref{eq:beta-L}), for FA-MM configuration. The error bars indicate the mean standard errors. Solid lines are numerical simulations for the FA-MM and SM-MM configurations }
 \end{center}
\end{figure}

In order to determine the mode matching between the fiber and the cavity modes, we use an FA as the input of the cavity and a MM fiber, with a laser machined and high reflectivity coated end facet, as the output. The FA has a designed waist size of 8.1 $\mu m$ at a distance of 230 $\mu m$ from the end facet of the assembly and a radius of curvature (ROC) of 700 $\mu m$. The MM fiber for the cavity output has a ROC of 540 $\mu m$. Using these parameters, the expected mode matching from the numerical simulation is better than 90\% for cavity lengths between 180 $\mu m$ and 520 $\mu m$, with a maximum of 99.9\% at 460 $\mu m$ (shown in Fig.\ref{fig:modematch}).
 
Employing a MM fiber at the output of the cavity is beneficial as  the entire cavity transmission spectrum can be observed without spatially filtering high order transverse modes of the cavity. As described in \cite{Bick2016}, the mode matching of a FFPC can not be determined by simultaneously observing the cavity's reflection and transmission since the reflected light at the cavity input is mode-filtered by the SM input fiber. Thus, utilizing the resonant dips in the reflection  would greatly overestimate the cavity coupling. Therefore in this work, only the transmission spectra are used to determine the mode matching. 

The ratio of the cavity transmission intensities of the fundamental TEM$_{\small{00}}$ mode ($T_{\small{00}}$) and of the modes of order $n,\,m$ ($T_{nm}$) defined by
\begin{equation}
 \beta = \frac{T_{\small{00}}}{\sum_{n,m}T_{nm}},
 \label{eq:beta}
\end{equation}
can be used to estimate the real mode matching efficiency $\eta_{\small{00}}$ of the fiber mode to the cavity's fundamental mode.
This approximation for the mode matching ($\beta = \eta_{\small{00}}$) is valid as long as the damping for all the cavity modes is the same.
The transmission intensity $T_{nm}$ is not only a function of the mode-matching efficiency $\eta_{nm}$ but also of the cavity finesse $\mathcal{F}_{nm}$ for the corresponding mode:
\begin{align}
 T_{nm} = \tilde{\eta}\,\eta_{nm}\,\mathcal{F}_{nm}^2 I_{in}, \label{eq:T}
\end{align}  
where $\tilde{\eta}$ includes the efficiencies independent of the mode orders, as well as the detection and propagation efficiencies, and $I_{in}$ is the intensity of the input beam. When the values of $\mathcal{F}_{nm}$ are the same irrespective of the mode order, $\beta = \eta_{\small{00}}$ is satisfied in Eq.~(\ref{eq:beta}) from the normalization condition $\sum_{n,m}\eta_{\small{nm}}= 1$.
Otherwise the contribution of the higher order modes may be underestimated, leading to an overestimation of the mode matching. In particular, clipping losses \cite{Hunger2010} can lead to an increased damping for the high order modes due to their larger spatial extension on the cavity mirrors.

To eliminate this effect on our measurements, we measure the finesses of all the modes visible in the transmission spectrum to confirm uniform damping of the modes. To this end, the cavity length, $L_{c}$, is measured using a calibrated microscope and the free spectral range (FSR) is calculated as FSR$=c/2L_{c}$. An electro-optic modulator is used to generate sidebands on the laser to act as a frequency calibration of the transmission spectrum to evaluate the cavity linewidth $\delta\nu$ \cite{Takahashi2014}. The finesse is then calculated by $\mathcal{F}\,=\,\mbox{FSR}/\delta\nu$.
\begin{figure}[bt]
 \includegraphics[width=\linewidth]{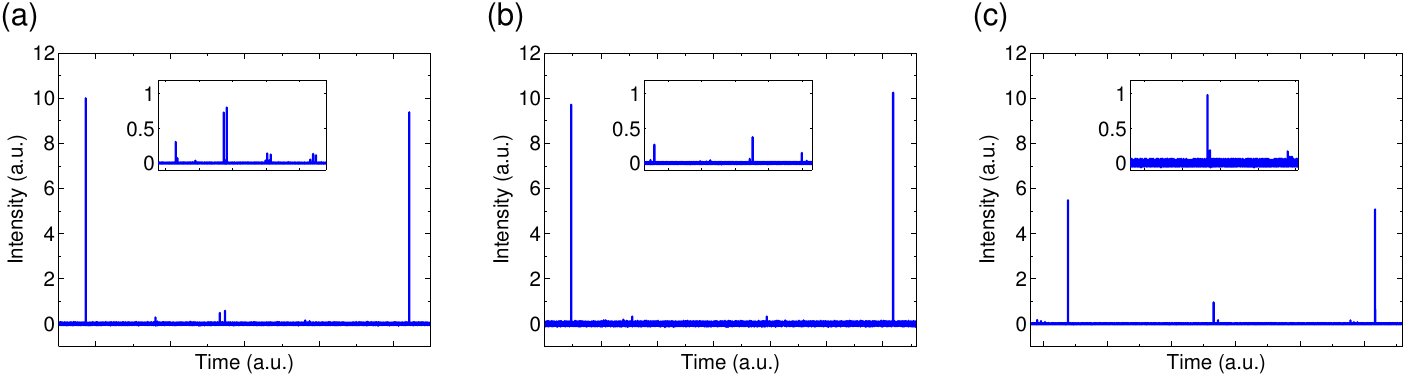}
 \caption{Transmission spectra at three different mean cavity lengths, $L_{c}$\,=\,(a) 132\,$\mu$m, (b) 426\,$\mu$m, (c) 536\, $\mu$m while the cavity is scanned by the piezo actuator over approximately one FSR. The inset spectra  magnify the middle parts of the scans to show higher order modes.}
 \label{fig:spect}
\end{figure}
\begin{figure}[bt]
  \begin{center}
  \includegraphics[scale=0.75]{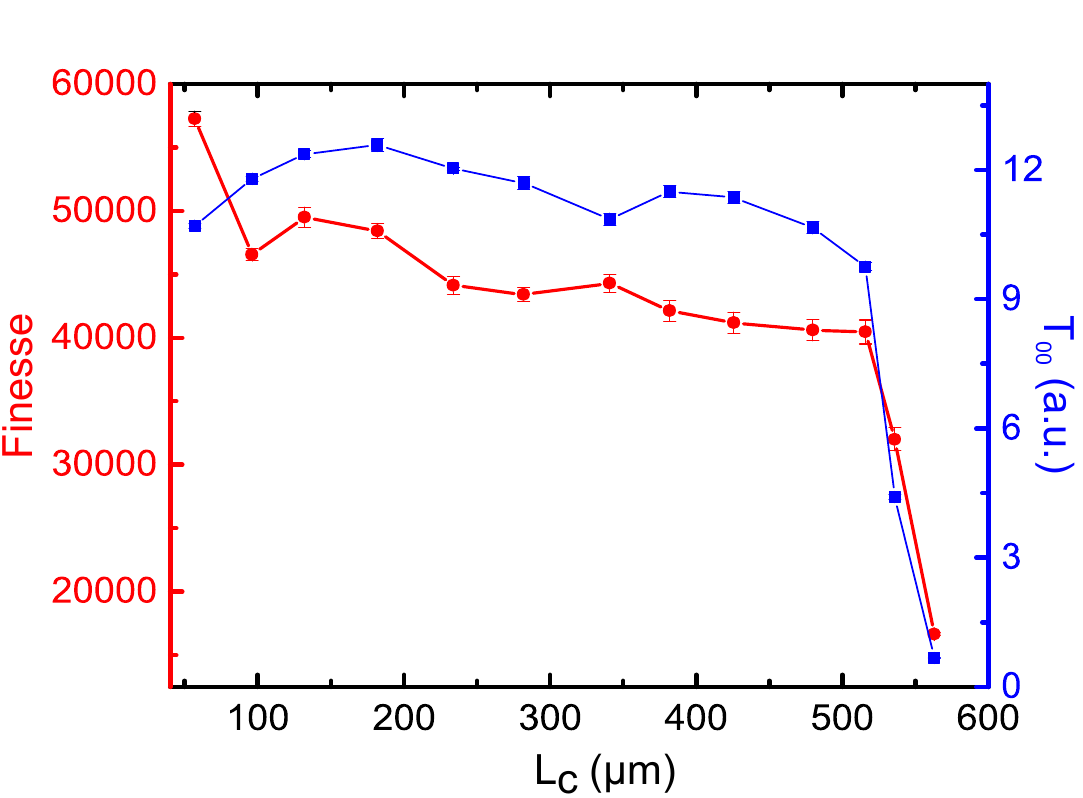}
  \caption{
Finesse (circle) and transmission intensities (square), for the fundamental mode, ($T_{\small{00}}$) as a function of cavity length ($L_{c}$) for the FA-MM configuration described in the text. An average of 20 traces is taken for each data point. The error bars indicate the mean standard errors.}
   \label{fig:fa_mm}
 \end{center}
\end{figure}

In Fig.~\ref{fig:spect} the transmission spectra at three different cavity lengths are displayed with the insets showing details of the higher order mode peaks. In addition, Fig.~\ref{fig:fa_mm} shows the finesse as well as the transmission intensity, $T_{\small{00}}$ of the fundamental mode as a function of the cavity length.
The finesse exhibits the typical behavior of a fiber cavity with MM fiber mirrors \cite{Brandstatter2013, Takahashi2014,Benedikter2015} with a plateau across most of the length range and a sharp drop at the edge of the stability limit. Similarly, $T_{\small{00}}$ is stable across most of the stability region of the cavity. However, the transmission spectra show a very different behavior with $\beta$ = 74\% at a cavity length of 132\,$\mu m$, 69\% at 536\,$\mu m$ and a 90\% at the cavity length of 426\,$\mu m$. At these three cavity lengths, we confirmed that the cavity finesse is identical within the measurement errors for all the modes visible in the spectrum, and hence, $\beta$ is a good measure for mode matching.  

The mode matching at a cavity length $L_c$ is related to the one at $L_c^{(0)}$ through
\begin{align}
 \eta_{00}(L_c) = \frac{T_{00}(L_c)}{T_{00}(L_c^{(0)})}\left(\frac{\mathcal{F}_{00}(L_c^{(0)})}{\mathcal{F}_{00}(L_c)}\right)^2\eta_{00}(L_c^{(0)}). \label{eq:beta-L}
\end{align}
Hence, from a single measurement of $\eta_{00}$ at a cavity length $L_c^{(0)}$, the mode matching efficiency at an arbitrary cavity length can be calculated by using only the transmission and finesse of the fundamental mode.     
Here we use the measurement at a cavity length of 132 $\mathrm{\mu m}$ ($= L_c^{(0)}$) as reference where $\beta \approx \eta_{00}$ holds, and calculate the estimated mode matching efficiencies over the entire range of the cavity length by using Eq.~(\ref{eq:beta-L}) and the data set in Fig.~\ref{fig:fa_mm}.  
As shown in Fig.\ref{fig:modematch}, the mode matching efficiencies calculated in this way fit well with the numerical prediction, constituting another strong evidence that our mode-matching assembly is working as expected. Fig.~\ref{fig:modematch} also shows a numerical simulation where the same cavity geometry is used but the FA is replaced with a SM fiber. One can see the substantial advantage of our FA for cavity lengths $> 150\,\mathrm{\mu m}$. 

\begin{figure}[ht]
  \begin{center}
    \includegraphics[width=0.8\linewidth]{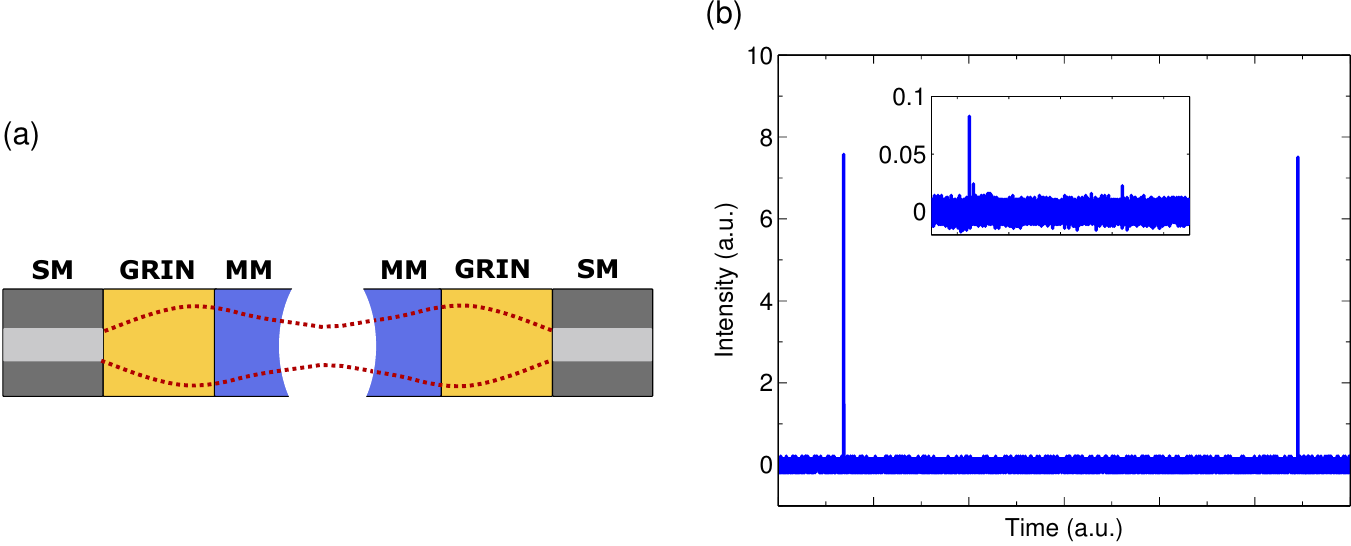}
   \caption{
   (a) Schematic of a FA-FA system where mode matching fiber assemblies are used as the input as well as the output of the FFPC. (b) Transmission spectrum of the FA-FA system as the cavity is scanned by a piezo actuator over approximately one FSR. The inset shows details of the higher order mode transmission. }
   \label{fig:modefilter}
 \end{center}
\end{figure}

Another interesting system is one where both sides of the cavity are constituted by the mode matching fiber assemblies (FA-FA system). This configuration implements a bi-directional input-output system (see Fig.~\ref{fig:modefilter}a). 
Fig.~\ref{fig:modefilter}b shows a transmission spectrum for such a system with the input side of the cavity being the FA described above. The output side of the cavity is formed by another FA with ROC of 630 $\mathrm{\mu m}$ and designed for a waist size of 8.4 $\mathrm{\mu m}$ at a distance of 230 $\mathrm{\mu m}$ from the end facet of assembly. The length of the cavity is adjusted for the optimal mode matching between the fiber and the cavity modes. 
Here we see, the intensity ratio $\beta$ of 98.6 \%  at a cavity length of 420 $\mu m$, which is a result of the mode filtering happening on both sides of the cavity. In other words, in this scenario, the transmission $T_{nm}$ is proportional to $\eta_{nm}^2$ instead of $\eta_{nm}$, resulting in enhanced suppression of the higher order peaks.
This value of $\beta$ in turn corresponds to $\eta_{00} = 89 \%$ on average for a single FA, which is consistent with the results in Fig.~\ref{fig:spect} at the same cavity length.   
 
In this letter we have demonstrated a versatile method to optimize the mode matching between input/output optical fibers and FFPCs. Using a combination of GRIN and MM fibers, the modes of almost any SM fiber can be matched to the mode of a FFPC. This enables the use of standard polarization maintaining fibers as well as specialty fibers such as spun fibers or photonic crystal fibers in conjunction with FFPCs.
With mode matching efficiencies of about 90\%, FFPCs can now be utilized for a wide range of applications where good coupling between the cavity an optical SM fibers is crucial.

\section{Acknowledgments}
We gratefully acknowledge support from EPSRC through the UK Quantum Technology
Hub: NQIT - Networked Quantum Information Technologies (EP/M013243/1) and
EP/J003670/1. G.K.G., H.T. and M.K. thank Qiming Wu for his help in the preparation of the fibers. 

\section{Competing financial interests}
The authors declare no competing financial interests.

\begin{thebibliography}{21}%
\makeatletter
\providecommand \@ifxundefined [1]{%
 \@ifx{#1\undefined}
}%
\providecommand \@ifnum [1]{%
 \ifnum #1\expandafter \@firstoftwo
 \else \expandafter \@secondoftwo
 \fi
}%
\providecommand \@ifx [1]{%
 \ifx #1\expandafter \@firstoftwo
 \else \expandafter \@secondoftwo
 \fi
}%
\providecommand \natexlab [1]{#1}%
\providecommand \enquote  [1]{``#1''}%
\providecommand \bibnamefont  [1]{#1}%
\providecommand \bibfnamefont [1]{#1}%
\providecommand \citenamefont [1]{#1}%
\providecommand \href@noop [0]{\@secondoftwo}%
\providecommand \href [0]{\begingroup \@sanitize@url \@href}%
\providecommand \@href[1]{\@@startlink{#1}\@@href}%
\providecommand \@@href[1]{\endgroup#1\@@endlink}%
\providecommand \@sanitize@url [0]{\catcode `\\12\catcode `\$12\catcode
  `\&12\catcode `\#12\catcode `\^12\catcode `\_12\catcode `\%12\relax}%
\providecommand \@@startlink[1]{}%
\providecommand \@@endlink[0]{}%
\providecommand \url  [0]{\begingroup\@sanitize@url \@url }%
\providecommand \@url [1]{\endgroup\@href {#1}{\urlprefix }}%
\providecommand \urlprefix  [0]{URL }%
\providecommand \Eprint [0]{\href }%
\providecommand \doibase [0]{http://dx.doi.org/}%
\providecommand \selectlanguage [0]{\@gobble}%
\providecommand \bibinfo  [0]{\@secondoftwo}%
\providecommand \bibfield  [0]{\@secondoftwo}%
\providecommand \translation [1]{[#1]}%
\providecommand \BibitemOpen [0]{}%
\providecommand \bibitemStop [0]{}%
\providecommand \bibitemNoStop [0]{.\EOS\space}%
\providecommand \EOS [0]{\spacefactor3000\relax}%
\providecommand \BibitemShut  [1]{\csname bibitem#1\endcsname}%
\let\auto@bib@innerbib\@empty
\bibitem [{\citenamefont {Hunger}\ \emph {et~al.}(2012)\citenamefont {Hunger},
  \citenamefont {Deutsch}, \citenamefont {Barbour}, \citenamefont {Warburton},\
  and\ \citenamefont {Reichel}}]{Hunger2012}%
  \BibitemOpen
  \bibfield  {author} {\bibinfo {author} {\bibfnamefont {D.}~\bibnamefont
  {Hunger}}, \bibinfo {author} {\bibfnamefont {C.}~\bibnamefont {Deutsch}},
  \bibinfo {author} {\bibfnamefont {R.}~\bibnamefont {Barbour}}, \bibinfo
  {author} {\bibfnamefont {R.}~\bibnamefont {Warburton}}, \ and\ \bibinfo
  {author} {\bibfnamefont {J.}~\bibnamefont {Reichel}},\ }\href
  {http://aip.scitation.org/doi/full/10.1063/1.3679721} {\bibfield  {journal}
  {\bibinfo  {journal} {AIP Adv.}\ }\textbf {\bibinfo {volume} {2}},\ \bibinfo
  {pages} {012119} (\bibinfo {year} {2012})}\BibitemShut {NoStop}%
\bibitem [{\citenamefont {Colombe}\ \emph {et~al.}(2007)\citenamefont
  {Colombe}, \citenamefont {Steinmetz}, \citenamefont {Dubois}, \citenamefont
  {Linke}, \citenamefont {Hunger},\ and\ \citenamefont
  {Reichel}}]{Colombe2007}%
  \BibitemOpen
  \bibfield  {author} {\bibinfo {author} {\bibfnamefont {Y.}~\bibnamefont
  {Colombe}}, \bibinfo {author} {\bibfnamefont {T.}~\bibnamefont {Steinmetz}},
  \bibinfo {author} {\bibfnamefont {G.}~\bibnamefont {Dubois}}, \bibinfo
  {author} {\bibfnamefont {F.}~\bibnamefont {Linke}}, \bibinfo {author}
  {\bibfnamefont {D.}~\bibnamefont {Hunger}}, \ and\ \bibinfo {author}
  {\bibfnamefont {J.}~\bibnamefont {Reichel}},\ }\href
  {http://www.nature.com/nature/journal/v450/n7167/full/nature06331.html}
  {\bibfield  {journal} {\bibinfo  {journal} {Nature}\ }\textbf {\bibinfo
  {volume} {450}},\ \bibinfo {pages} {272–276} (\bibinfo {year}
  {2007})}\BibitemShut {NoStop}%
\bibitem [{\citenamefont {Steiner}\ \emph {et~al.}(2013)\citenamefont
  {Steiner}, \citenamefont {Meyer}, \citenamefont {Deutsch}, \citenamefont
  {Reichel},\ and\ \citenamefont {K\"ohl}}]{Steiner2013}%
  \BibitemOpen
  \bibfield  {author} {\bibinfo {author} {\bibfnamefont {M.}~\bibnamefont
  {Steiner}}, \bibinfo {author} {\bibfnamefont {H.}~\bibnamefont {Meyer}},
  \bibinfo {author} {\bibfnamefont {C.}~\bibnamefont {Deutsch}}, \bibinfo
  {author} {\bibfnamefont {J.}~\bibnamefont {Reichel}}, \ and\ \bibinfo
  {author} {\bibfnamefont {M.}~\bibnamefont {K\"ohl}},\ }\href
  {http://link.aps.org/doi/10.1103/PhysRevLett.110.043003} {\bibfield
  {journal} {\bibinfo  {journal} {Phys. Rev. Lett.}\ }\textbf {\bibinfo
  {volume} {110}},\ \bibinfo {pages} {043003} (\bibinfo {year}
  {2013})}\BibitemShut {NoStop}%
\bibitem [{\citenamefont {Toninelli}\ \emph {et~al.}(2010)\citenamefont
  {Toninelli}, \citenamefont {Delley}, \citenamefont {St\"oferle},
  \citenamefont {Renn}, \citenamefont {G\"otzinger},\ and\ \citenamefont
  {Sandoghdar}}]{Tonielli2010}%
  \BibitemOpen
  \bibfield  {author} {\bibinfo {author} {\bibfnamefont {C.}~\bibnamefont
  {Toninelli}}, \bibinfo {author} {\bibfnamefont {Y.}~\bibnamefont {Delley}},
  \bibinfo {author} {\bibfnamefont {T.}~\bibnamefont {St\"oferle}}, \bibinfo
  {author} {\bibfnamefont {A.}~\bibnamefont {Renn}}, \bibinfo {author}
  {\bibfnamefont {S.}~\bibnamefont {G\"otzinger}}, \ and\ \bibinfo {author}
  {\bibfnamefont {V.}~\bibnamefont {Sandoghdar}},\ }\href
  {http://aip.scitation.org/doi/10.1063/1.3456559} {\bibfield  {journal}
  {\bibinfo  {journal} {Appl. Phys. Lett.}\ }\textbf {\bibinfo {volume} {97}},\
  \bibinfo {pages} {021107} (\bibinfo {year} {2010})}\BibitemShut {NoStop}%
\bibitem [{\citenamefont {Albrecht}\ \emph {et~al.}(2013)\citenamefont
  {Albrecht}, \citenamefont {Bommer}, \citenamefont {Deutsch}, \citenamefont
  {Reichel},\ and\ \citenamefont {Becher}}]{Albrecht2013}%
  \BibitemOpen
  \bibfield  {author} {\bibinfo {author} {\bibfnamefont {R.}~\bibnamefont
  {Albrecht}}, \bibinfo {author} {\bibfnamefont {A.}~\bibnamefont {Bommer}},
  \bibinfo {author} {\bibfnamefont {C.}~\bibnamefont {Deutsch}}, \bibinfo
  {author} {\bibfnamefont {J.}~\bibnamefont {Reichel}}, \ and\ \bibinfo
  {author} {\bibfnamefont {C.}~\bibnamefont {Becher}},\ }\href
  {http://link.aps.org/doi/10.1103/PhysRevLett.110.243602} {\bibfield
  {journal} {\bibinfo  {journal} {Phys. Rev. Lett.}\ }\textbf {\bibinfo
  {volume} {110}},\ \bibinfo {pages} {243602} (\bibinfo {year}
  {2013})}\BibitemShut {NoStop}%
\bibitem [{\citenamefont {Miguel-S\'anchez}\ \emph {et~al.}(2013)\citenamefont
  {Miguel-S\'anchez}, \citenamefont {Reinhard}, \citenamefont {Togan},
  \citenamefont {Volz}, \citenamefont {Imamo\'glu}, \citenamefont {Besga},
  \citenamefont {Reichel},\ and\ \citenamefont {Est\'eve}}]{Miguel2013}%
  \BibitemOpen
  \bibfield  {author} {\bibinfo {author} {\bibfnamefont {M.}~\bibnamefont
  {Miguel-S\'anchez}}, \bibinfo {author} {\bibfnamefont {A.}~\bibnamefont
  {Reinhard}}, \bibinfo {author} {\bibfnamefont {E.}~\bibnamefont {Togan}},
  \bibinfo {author} {\bibfnamefont {T.}~\bibnamefont {Volz}}, \bibinfo {author}
  {\bibfnamefont {A.}~\bibnamefont {Imamo\'glu}}, \bibinfo {author}
  {\bibfnamefont {B.}~\bibnamefont {Besga}}, \bibinfo {author} {\bibfnamefont
  {J.}~\bibnamefont {Reichel}}, \ and\ \bibinfo {author} {\bibfnamefont
  {J.}~\bibnamefont {Est\'eve}},\ }\href
  {http://iopscience.iop.org/article/10.1088/1367-2630/15/4/045002/meta}
  {\bibfield  {journal} {\bibinfo  {journal} {New J. Phys.}\ }\textbf {\bibinfo
  {volume} {15}},\ \bibinfo {pages} {045002} (\bibinfo {year}
  {2013})}\BibitemShut {NoStop}%
\bibitem [{\citenamefont {Besga}\ \emph {et~al.}(2015)\citenamefont {Besga},
  \citenamefont {Vaneph}, \citenamefont {Reichel}, \citenamefont {Est\'eve},
  \citenamefont {Reinhard}, \citenamefont {Miguel-S\'anchez}, \citenamefont
  {Imamo\'glu},\ and\ \citenamefont {Volz}}]{Besga2015}%
  \BibitemOpen
  \bibfield  {author} {\bibinfo {author} {\bibfnamefont {B.}~\bibnamefont
  {Besga}}, \bibinfo {author} {\bibfnamefont {C.}~\bibnamefont {Vaneph}},
  \bibinfo {author} {\bibfnamefont {J.}~\bibnamefont {Reichel}}, \bibinfo
  {author} {\bibfnamefont {J.}~\bibnamefont {Est\'eve}}, \bibinfo {author}
  {\bibfnamefont {A.}~\bibnamefont {Reinhard}}, \bibinfo {author}
  {\bibfnamefont {J.}~\bibnamefont {Miguel-S\'anchez}}, \bibinfo {author}
  {\bibfnamefont {A.}~\bibnamefont {Imamo\'glu}}, \ and\ \bibinfo {author}
  {\bibfnamefont {T.}~\bibnamefont {Volz}},\ }\href
  {http://journals.aps.org/prapplied/abstract/10.1103/PhysRevApplied.3.014008}
  {\bibfield  {journal} {\bibinfo  {journal} {Phys. Rev. Applied}\ }\textbf
  {\bibinfo {volume} {3}},\ \bibinfo {pages} {014008} (\bibinfo {year}
  {2015})}\BibitemShut {NoStop}%
\bibitem [{\citenamefont {Flowers-Jacobs}\ \emph {et~al.}(2016)\citenamefont
  {Flowers-Jacobs}, \citenamefont {Hoch}, \citenamefont {Sankey}, \citenamefont
  {Kashkanova}, \citenamefont {Jayich}, \citenamefont {Deutsch}, \citenamefont
  {Reichel},\ and\ \citenamefont {Harris}}]{Flowers2016}%
  \BibitemOpen
  \bibfield  {author} {\bibinfo {author} {\bibfnamefont {N.}~\bibnamefont
  {Flowers-Jacobs}}, \bibinfo {author} {\bibfnamefont {S.}~\bibnamefont
  {Hoch}}, \bibinfo {author} {\bibfnamefont {J.}~\bibnamefont {Sankey}},
  \bibinfo {author} {\bibfnamefont {A.}~\bibnamefont {Kashkanova}}, \bibinfo
  {author} {\bibfnamefont {A.}~\bibnamefont {Jayich}}, \bibinfo {author}
  {\bibfnamefont {C.}~\bibnamefont {Deutsch}}, \bibinfo {author} {\bibfnamefont
  {J.}~\bibnamefont {Reichel}}, \ and\ \bibinfo {author} {\bibfnamefont
  {J.}~\bibnamefont {Harris}},\ }\href
  {http://aip.scitation.org/doi/10.1063/1.4768779} {\bibfield  {journal}
  {\bibinfo  {journal} {Appl. Phys. Lett.}\ }\textbf {\bibinfo {volume}
  {101}},\ \bibinfo {pages} {221109} (\bibinfo {year} {2016})}\BibitemShut
  {NoStop}%
\bibitem [{\citenamefont {Miller}\ and\ \citenamefont
  {Janniello}(1990)}]{Miller1990}%
  \BibitemOpen
  \bibfield  {author} {\bibinfo {author} {\bibfnamefont {C.}~\bibnamefont
  {Miller}}\ and\ \bibinfo {author} {\bibfnamefont {F.}~\bibnamefont
  {Janniello}},\ }\href {http://ieeexplore.ieee.org/document/59633/} {\bibfield
   {journal} {\bibinfo  {journal} {Electron. Lett.}\ }\textbf {\bibinfo
  {volume} {26}},\ \bibinfo {pages} {2122} (\bibinfo {year}
  {1990})}\BibitemShut {NoStop}%
\bibitem [{\citenamefont {Mader}\ \emph {et~al.}(2015)\citenamefont {Mader},
  \citenamefont {Reichel}, \citenamefont {H\"ansch},\ and\ \citenamefont
  {Hunger}}]{Mader2015}%
  \BibitemOpen
  \bibfield  {author} {\bibinfo {author} {\bibfnamefont {M.}~\bibnamefont
  {Mader}}, \bibinfo {author} {\bibfnamefont {J.}~\bibnamefont {Reichel}},
  \bibinfo {author} {\bibfnamefont {T.~W.}\ \bibnamefont {H\"ansch}}, \ and\
  \bibinfo {author} {\bibfnamefont {D.}~\bibnamefont {Hunger}},\ }\href
  {http://www.nature.com/articles/ncomms8249} {\bibfield  {journal} {\bibinfo
  {journal} {Nature Comm.}\ }\textbf {\bibinfo {volume} {6}},\ \bibinfo {pages}
  {7249} (\bibinfo {year} {2015})}\BibitemShut {NoStop}%
\bibitem [{\citenamefont {Petrak}\ \emph {et~al.}(2014)\citenamefont {Petrak},
  \citenamefont {Djeu},\ and\ \citenamefont {Muller}}]{Petrak2014}%
  \BibitemOpen
  \bibfield  {author} {\bibinfo {author} {\bibfnamefont {B.}~\bibnamefont
  {Petrak}}, \bibinfo {author} {\bibfnamefont {N.}~\bibnamefont {Djeu}}, \ and\
  \bibinfo {author} {\bibfnamefont {A.}~\bibnamefont {Muller}},\ }\href
  {https://journals.aps.org/pra/abstract/10.1103/PhysRevA.89.023811} {\bibfield
   {journal} {\bibinfo  {journal} {Phys. Rev. A}\ }\textbf {\bibinfo {volume}
  {89}},\ \bibinfo {pages} {023811} (\bibinfo {year} {2014})}\BibitemShut
  {NoStop}%
\bibitem [{\citenamefont {Ni}\ \emph {et~al.}(2016)\citenamefont {Ni},
  \citenamefont {Fu},\ and\ \citenamefont {Zhao}}]{Ni2016}%
  \BibitemOpen
  \bibfield  {author} {\bibinfo {author} {\bibfnamefont {X.}~\bibnamefont
  {Ni}}, \bibinfo {author} {\bibfnamefont {S.}~\bibnamefont {Fu}}, \ and\
  \bibinfo {author} {\bibfnamefont {Z.}~\bibnamefont {Zhao}},\ }\href {\doibase
  10.1109/JPHOT.2016.2566448} {\bibfield  {journal} {\bibinfo  {journal} {IEEE
  Photonics Journal}\ }\textbf {\bibinfo {volume} {8}},\ \bibinfo {pages} {1}
  (\bibinfo {year} {2016})}\BibitemShut {NoStop}%
\bibitem [{\citenamefont {Takahashi}\ \emph {et~al.}(2014)\citenamefont
  {Takahashi}, \citenamefont {Morphew}, \citenamefont {Oru\^cevi\'c},
  \citenamefont {Noguchi}, \citenamefont {Kassa},\ and\ \citenamefont
  {Keller}}]{Takahashi2014}%
  \BibitemOpen
  \bibfield  {author} {\bibinfo {author} {\bibfnamefont {H.}~\bibnamefont
  {Takahashi}}, \bibinfo {author} {\bibfnamefont {J.}~\bibnamefont {Morphew}},
  \bibinfo {author} {\bibfnamefont {F.}~\bibnamefont {Oru\^cevi\'c}}, \bibinfo
  {author} {\bibfnamefont {A.}~\bibnamefont {Noguchi}}, \bibinfo {author}
  {\bibfnamefont {E.}~\bibnamefont {Kassa}}, \ and\ \bibinfo {author}
  {\bibfnamefont {M.}~\bibnamefont {Keller}},\ }\href
  {https://www.osapublishing.org/oe/abstract.cfm?uri=oe-22-25-31317} {\bibfield
   {journal} {\bibinfo  {journal} {Opt. Express}\ }\textbf {\bibinfo {volume}
  {22}},\ \bibinfo {pages} {31317} (\bibinfo {year} {2014})}\BibitemShut
  {NoStop}%
\bibitem [{\citenamefont {Ott}\ \emph {et~al.}(2016)\citenamefont {Ott},
  \citenamefont {Garcia},\ and\ \citenamefont {Kohlhaas}}]{Ott2016}%
  \BibitemOpen
  \bibfield  {author} {\bibinfo {author} {\bibfnamefont {K.}~\bibnamefont
  {Ott}}, \bibinfo {author} {\bibfnamefont {S.}~\bibnamefont {Garcia}}, \ and\
  \bibinfo {author} {\bibfnamefont {R.}~\bibnamefont {Kohlhaas}},\ }\href
  {\doibase 10.1364/OE.24.009839} {\bibfield  {journal} {\bibinfo  {journal}
  {Opt. Expresss}\ }\textbf {\bibinfo {volume} {24}},\ \bibinfo {pages} {9839}
  (\bibinfo {year} {2016})}\BibitemShut {NoStop}%
\bibitem [{\citenamefont {Bick}\ \emph {et~al.}(2016)\citenamefont {Bick},
  \citenamefont {Staarmann}, \citenamefont {Christoph}, \citenamefont
  {Hellmig}, \citenamefont {Heinze}, \citenamefont {Sengstock},\ and\
  \citenamefont {Becker}}]{Bick2016}%
  \BibitemOpen
  \bibfield  {author} {\bibinfo {author} {\bibfnamefont {A.}~\bibnamefont
  {Bick}}, \bibinfo {author} {\bibfnamefont {C.}~\bibnamefont {Staarmann}},
  \bibinfo {author} {\bibfnamefont {P.}~\bibnamefont {Christoph}}, \bibinfo
  {author} {\bibfnamefont {O.}~\bibnamefont {Hellmig}}, \bibinfo {author}
  {\bibfnamefont {J.}~\bibnamefont {Heinze}}, \bibinfo {author} {\bibfnamefont
  {K.}~\bibnamefont {Sengstock}}, \ and\ \bibinfo {author} {\bibfnamefont
  {C.}~\bibnamefont {Becker}},\ }\href
  {http://aip.scitation.org/doi/full/10.1063/1.4939046} {\bibfield  {journal}
  {\bibinfo  {journal} {Review of Scientific Instruments}\ }\textbf {\bibinfo
  {volume} {87}} (\bibinfo {year} {2016})}\BibitemShut {NoStop}%
\bibitem [{\citenamefont {Gallego}\ \emph {et~al.}(2016)\citenamefont
  {Gallego}, \citenamefont {Ghosh}, \citenamefont {Alavi}, \citenamefont {Alt},
  \citenamefont {Martinez-Dorantes}, \citenamefont {Meschede},\ and\
  \citenamefont {Ratschbacher}}]{Gallego2016}%
  \BibitemOpen
  \bibfield  {author} {\bibinfo {author} {\bibfnamefont {J.}~\bibnamefont
  {Gallego}}, \bibinfo {author} {\bibfnamefont {S.}~\bibnamefont {Ghosh}},
  \bibinfo {author} {\bibfnamefont {S.~K.}\ \bibnamefont {Alavi}}, \bibinfo
  {author} {\bibfnamefont {W.}~\bibnamefont {Alt}}, \bibinfo {author}
  {\bibfnamefont {M.}~\bibnamefont {Martinez-Dorantes}}, \bibinfo {author}
  {\bibfnamefont {D.}~\bibnamefont {Meschede}}, \ and\ \bibinfo {author}
  {\bibfnamefont {L.}~\bibnamefont {Ratschbacher}},\ }\href
  {https://link.springer.com/article/10.1007/s00340-015-6281-z} {\bibfield
  {journal} {\bibinfo  {journal} {Applied Physics B}\ }\textbf {\bibinfo
  {volume} {122}},\ \bibinfo {pages} {47} (\bibinfo {year} {2016})}\BibitemShut
  {NoStop}%
\bibitem [{\citenamefont {Benedikter}\ \emph {et~al.}(2015)\citenamefont
  {Benedikter}, \citenamefont {H\"ummer}, \citenamefont {Mader}, \citenamefont
  {Schlederer}, \citenamefont {Reichel}, \citenamefont {H\"ansch},\ and\
  \citenamefont {Hunger}}]{Benedikter2015}%
  \BibitemOpen
  \bibfield  {author} {\bibinfo {author} {\bibfnamefont {J.}~\bibnamefont
  {Benedikter}}, \bibinfo {author} {\bibfnamefont {T.}~\bibnamefont
  {H\"ummer}}, \bibinfo {author} {\bibfnamefont {M.}~\bibnamefont {Mader}},
  \bibinfo {author} {\bibfnamefont {B.}~\bibnamefont {Schlederer}}, \bibinfo
  {author} {\bibfnamefont {J.}~\bibnamefont {Reichel}}, \bibinfo {author}
  {\bibfnamefont {T.}~\bibnamefont {H\"ansch}}, \ and\ \bibinfo {author}
  {\bibfnamefont {D.}~\bibnamefont {Hunger}},\ }\href
  {http://iopscience.iop.org/article/10.1088/1367-2630/17/5/053051} {\bibfield
  {journal} {\bibinfo  {journal} {New J. Phys.}\ }\textbf {\bibinfo {volume}
  {17}},\ \bibinfo {pages} {053051} (\bibinfo {year} {2015})}\BibitemShut
  {NoStop}%
\bibitem [{\citenamefont {Podoliak}\ \emph {et~al.}(2017)\citenamefont
  {Podoliak}, \citenamefont {Takahashi}, \citenamefont {Keller},\ and\
  \citenamefont {Horak}}]{Podoliak2017}%
  \BibitemOpen
  \bibfield  {author} {\bibinfo {author} {\bibfnamefont {N.}~\bibnamefont
  {Podoliak}}, \bibinfo {author} {\bibfnamefont {H.}~\bibnamefont {Takahashi}},
  \bibinfo {author} {\bibfnamefont {M.}~\bibnamefont {Keller}}, \ and\ \bibinfo
  {author} {\bibfnamefont {P.}~\bibnamefont {Horak}},\ }\href
  {https://doi.org/10.1088/1361-6455/aa640a} {\bibfield  {journal} {\bibinfo
  {journal} {J. Phys. B: At. Mol. Opt. Phys. in press}\ } (\bibinfo {year}
  {2017})}\BibitemShut {NoStop}%
\bibitem [{\citenamefont {Wang}\ \emph {et~al.}(2014)\citenamefont {Wang},
  \citenamefont {Zhang}, \citenamefont {Bi}, \citenamefont {Xia},\ and\
  \citenamefont {Xu}}]{Wang2014}%
  \BibitemOpen
  \bibfield  {author} {\bibinfo {author} {\bibfnamefont {C.}~\bibnamefont
  {Wang}}, \bibinfo {author} {\bibfnamefont {F.}~\bibnamefont {Zhang}},
  \bibinfo {author} {\bibfnamefont {S.-B.}\ \bibnamefont {Bi}}, \bibinfo
  {author} {\bibfnamefont {X.-Q.}\ \bibnamefont {Xia}}, \ and\ \bibinfo
  {author} {\bibfnamefont {T.-T.}\ \bibnamefont {Xu}},\ }\href {\doibase
  10.1007/s40436-014-0089-7} {\bibfield  {journal} {\bibinfo  {journal}
  {Advances in Manufacturing}\ }\textbf {\bibinfo {volume} {2}},\ \bibinfo
  {pages} {327} (\bibinfo {year} {2014})}\BibitemShut {NoStop}%
\bibitem [{\citenamefont {Hunger}\ \emph {et~al.}(2010)\citenamefont {Hunger},
  \citenamefont {Steinmetz}, \citenamefont {Colombe}, \citenamefont {Deutsch},
  \citenamefont {H\"{a}nsch},\ and\ \citenamefont {Reichel}}]{Hunger2010}%
  \BibitemOpen
  \bibfield  {author} {\bibinfo {author} {\bibfnamefont {D.}~\bibnamefont
  {Hunger}}, \bibinfo {author} {\bibfnamefont {T.}~\bibnamefont {Steinmetz}},
  \bibinfo {author} {\bibfnamefont {Y.}~\bibnamefont {Colombe}}, \bibinfo
  {author} {\bibfnamefont {C.}~\bibnamefont {Deutsch}}, \bibinfo {author}
  {\bibfnamefont {T.~W.}\ \bibnamefont {H\"{a}nsch}}, \ and\ \bibinfo {author}
  {\bibfnamefont {J.}~\bibnamefont {Reichel}},\ }\href {\doibase
  10.1088/1367-2630/12/6/065038} {\bibfield  {journal} {\bibinfo  {journal}
  {New Journal of Physics}\ }\textbf {\bibinfo {volume} {12}},\ \bibinfo
  {pages} {065038} (\bibinfo {year} {2010})}\BibitemShut {NoStop}%
\bibitem [{\citenamefont {Brandst\"atter}\ \emph {et~al.}(2013)\citenamefont
  {Brandst\"atter}, \citenamefont {McClung}, \citenamefont {Schuppert},
  \citenamefont {Casabone}, \citenamefont {Friebe}, \citenamefont {Stute},
  \citenamefont {Schmidt}, \citenamefont {Deutsch}, \citenamefont {Reiche},
  \citenamefont {Blatt},\ and\ \citenamefont {Northup}}]{Brandstatter2013}%
  \BibitemOpen
  \bibfield  {author} {\bibinfo {author} {\bibfnamefont {B.}~\bibnamefont
  {Brandst\"atter}}, \bibinfo {author} {\bibfnamefont {A.}~\bibnamefont
  {McClung}}, \bibinfo {author} {\bibfnamefont {K.}~\bibnamefont {Schuppert}},
  \bibinfo {author} {\bibfnamefont {B.}~\bibnamefont {Casabone}}, \bibinfo
  {author} {\bibfnamefont {K.}~\bibnamefont {Friebe}}, \bibinfo {author}
  {\bibfnamefont {A.}~\bibnamefont {Stute}}, \bibinfo {author} {\bibfnamefont
  {P.}~\bibnamefont {Schmidt}}, \bibinfo {author} {\bibfnamefont
  {C.}~\bibnamefont {Deutsch}}, \bibinfo {author} {\bibfnamefont
  {J.}~\bibnamefont {Reiche}}, \bibinfo {author} {\bibfnamefont
  {R.}~\bibnamefont {Blatt}}, \ and\ \bibinfo {author} {\bibfnamefont
  {T.}~\bibnamefont {Northup}},\ }\href
  {http://aip.scitation.org/doi/10.1063/1.4838696} {\bibfield  {journal}
  {\bibinfo  {journal} {Rev. Sci. Instrum.}\ }\textbf {\bibinfo {volume}
  {84}},\ \bibinfo {pages} {123104} (\bibinfo {year} {2013})}\BibitemShut
  {NoStop}%
\end{thebibliography}
 \end{document}